\documentclass[iop]{emulateapj}
\usepackage{natbib}
\usepackage{epsfig}
\usepackage{graphicx}
\usepackage{epstopdf}
\newcommand\icarus{{Icarus}}

\newcommand{\rau}{r_{\rm{AU}}}

\newcommand{\unitkappa}{{\rm cm^{2}\:g^{-1}}}
\newcommand{\sigmasb}{\sigma_{\rm{SB}}}

\newcommand{\gccc}{\rm{g\:cm^{-3}}}
\newcommand{\kepler}{{\it{Kepler}}}
\newcommand{\mpone}{M_{p,1}}
\newcommand{\rpone}{R_{p,1}}
\newcommand{\rearth}{{R_\oplus}}
\newcommand{\days}{{\rm{day}}}
\newcommand{\insitu}{{\it{in situ}}}
\newcommand{\au}{{\rm{AU}}}
\newcommand{\phidzib}{{\phi_{\rm{DZIB}}}}
\newcommand{\unitmdot}{{M_\odot \rm{yr}^{-1}}}
\newcommand{\msun}{{M_\odot}}

\newcommand{\mearth}{{M_\oplus}}

\newcommand     \K              {\,{\rm K}}

\newcommand     \yr     {\,{\rm yr}}

\newcommand{\smyr}{{ M_\odot\:\rm yr^{-1}}}

\newcommand{\beq}{\begin{equation}}
\newcommand{\eeq}{\end{equation}}
\newcommand{\beqa}{\begin{eqnarray}}
\newcommand{\eeqa}{\end{eqnarray}}

\newlength{\figwidth}
\shorttitle{Innermost Planets from IOPF}
\shortauthors{Chatterjee \& Tan}

\begin{document}
\title{Vulcan Planets: Inside-Out Formation of the Innermost Super-Earths}

\author{Sourav Chatterjee}
\affil{Center for Interdisciplinary Exploration and Research in Astrophysics (CIERA)\\
Physics \& Astronomy, Northwestern University, Evanston, IL 60208, USA\\
sourav.chatterjee@northwestern.edu}
\author{Jonathan C. Tan}
\affil{Departments of Astronomy \& Physics, University of Florida, Gainesville, FL 32611, USA\\jt@astro.ufl.edu}

\begin{abstract}
The compact multi-transiting systems discovered by \kepler\ challenge
traditional planet formation theories. These fall into two
broad classes: (1) formation further out followed by migration; (2)
formation {\it in situ} from a disk of 
gas and planetesimals. In the former, an abundance of resonant chains
is expected, which the \kepler\ data do not support. In the latter,
required disk mass surface densities
may be too high.  
A recently proposed mechanism hypothesizes that planets form {\it in
  situ} at the pressure trap associated with the dead-zone inner
boundary (DZIB) where radially drifting ``pebbles'' accumulate. This
scenario predicts planet masses ($M_p$) are set by the gap-opening
process that then leads to DZIB retreat, followed by sequential,
inside-out planet formation (IOPF). For typical disk accretion rates,
IOPF predictions for $M_p$, $M_p$ versus orbital radius $r$, and
planet-planet separations
are consistent with observed systems.
Here we investigate the IOPF prediction for how the masses, $\mpone$,
of the innermost 
(``Vulcan") planets vary with $r$. We show that for
fiducial parameters, $\mpone\simeq5.0(r/{\rm{0.1\:AU}})\:\mearth$,
independent of the disk's accretion rate at time of planet formation.
Then, using Monte Carlo sampling of a population of these innermost
planets, we test this predicted scaling against observed planet
properties, allowing for intrinsic dispersions in planetary densities
and {\it{Kepler's}} observational biases. These effects lead to a
slightly shallower relation $\mpone\propto{r}^{0.9\pm0.2}$, which is
consistent with $\mpone\propto{r}^{0.7\pm0.2}$ of the observed
Vulcans. The normalization of the relation constrains the gap-opening
process, favoring relatively low viscosities in the inner dead zone.
\end{abstract}

\section{Introduction}\label{S:intro}

A surprising discovery of NASA's \kepler\ mission is the existence of
multi-transiting planetary systems with tightly-packed inner planets
(STIPs): typically $3$--$5$-planet systems with radii
$\sim1$--$10\:\rearth$ in short-period ($1$--$100\:\days$) orbits
\citep{2012ApJ...761...92F}. Planet-planet scattering followed by
tidal circularization is unlikely to produce the observed low
dispersion ($\lesssim3^\circ$) in their mutual orbital inclinations
\citep[e.g.,][]{1996Sci...274..954R,2008ApJ...686..580C,2011ApJ...742...72N}.

Formation further out followed by inward, disk-mediated migration
\citep{2012ARA&A..50..211K,2013A&A...553L...2C,2014arXiv1407.6011C}
has been proposed. However, migration scenarios
may produce planetary orbits that are trapped near low-order
mean motion resonances (MMR). Such
orbits are not particularly common among the \kepler\ planet
candidates (KPCs). It has been argued that lower-mass planets,
  like KPCs, may not be efficiently trapped in resonance chains
  \citep{2012Icar..221..624M,2013ApJ...778....7B,2014AJ....147...32G}.
Other mechanisms, operating long after formation, may also move
planets out of resonance
\citep{2011CeMDA.111...83P,2012ApJ...756L..11L,2012MNRAS.427L..21R,2013AJ....145....1B,2013ApJ...770...24P,2014arXiv1406.0521C}.

{\it In situ} formation has also been proposed
\citep{2013MNRAS.431.3444C,2012ApJ...751..158H,2013ApJ...775...53H}.
Standard \insitu\ formation models face challenges of concentrating
the required large mass of solids extremely near the star, needing
disks $\geq20\times$ more massive than the minimum mass solar nebula
and widely varying density profiles to explain observed STIPs. Such
disks may not be compatible with standard viscous accretion disk
theory and a large fraction of them may not remain stable under
self-gravity for reasonable gas-to-dust ratios
\citep[][]{2014MNRAS.440L..11R,2014ApJ...795L..15S}.

Recently \citet[][henceforth CT14]{2014ApJ...780...53C} proposed an
alternative mechanism: ``Inside-Out Planet Formation" (IOPF), which
alleviates some of the above problems. In a typical, steadily
accreting disk, macroscopic, $\sim$cm-sized ``pebbles" formed from
dust grain coagulation should undergo rapid inward radial drift and
become trapped at the global pressure maxima expected at the dead-zone
inner boundary (DZIB), where the ionization fraction set by thermal
ionization of alkali metals drops below the critical value needed for
the magneto-rotational instability (MRI) to operate. A ring forms with
enhanced density of solids, promoting planet formation, perhaps first
via gravitational instability to form $\sim$moon-size objects. These
may then mutually collide to form a single dominant planet, which can
also grow by continued accretion of pebbles. Growth stops and planet
mass is set when the planet is massive enough to open a gap in the
disk leading to retreat of the DZIB and its associated pressure
maximum, and thus truncation of the supply of pebbles. This scenario
naturally alleviates challenges of solid enhancement near the star
since the pebble supply zone can be $\gtrsim10\:\au$
\citep{2014arXiv1410.5819H}. For typical disk accretion rates and
viscosities, predicted values of $M_p$, $M_p$-$r$ scalings for
individual systems, and planet-planet separations are consistent with
observed systems.

Here we focus on the innermost (``Vulcan'') planet mass, $\mpone$,
versus orbital radius, $r$, relation that naturally follows from IOPF
theory and test whether observed systems support this scaling law. 
\S\ref{S:derivation} derives the theoretical $\mpone$-$r$ relation.
\S\ref{S:obs} summarizes relevant observed properties of KPCs
allowing \S\ref{S:compare} to compare theory with observation.
\S\ref{S:conclude} concludes.

%
%
\begin{figure*}
\begin{center}
\plotone{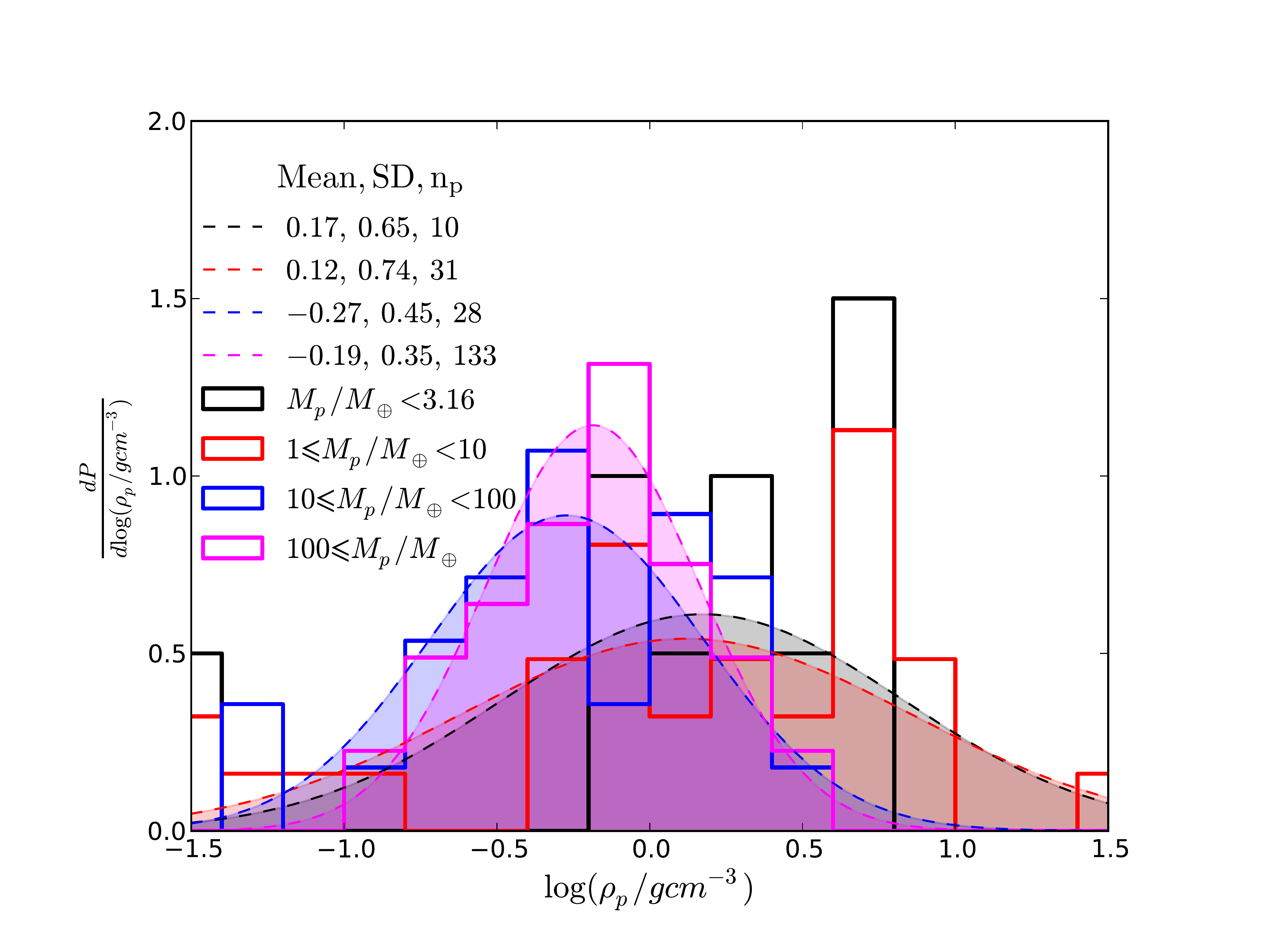}
\caption[$\rho_p$ distribution]{
Probability distribution functions for $\log(\rho_p/\gccc)$ of
observed planets with known $M_p$ and $R_p$. Full set is divided
into four groups contained in equal logarithmic bins of $M_p/\mearth$,
$0.1$--$1$ (black), $1$--$10$ (red), $10$--$10^2$ (blue),
$10^2$--$10^3$ (magenta).  Since no exoplanets with known $M_p$ and
$R_p$ populate the first group $0.1\leq(M_p/M_\oplus)<1$, we estimate
the $\rho_p$-distribution by taking into account all planets with
$M_p/M_\oplus\leqslant3.16$.
Solid histograms and filled dashed lines show distributions of actual
data and best-fit lognormals, respectively. Legend shows mean and
standard deviation for each distribution, and number of planets in
each group (left to right, respectively).
}
\label{fig:rho}
\end{center}
\end{figure*}
%
%
\section{Innermost Planet Mass vs Orbital Radius Relation}
\label{S:derivation}
IOPF theory predicts that position of formation of the innermost
planet is determined by DZIB location, first set by thermal ionzation
of alkali metals at disk midplane temperatures
$T\simeq1200\:\K$. Predicting this location is expected to be
relatively simple compared to locations of subsequent planet
formation, which depend on extent of DZIB retreat, which depends on
reduction of disk gas column density caused by presence of the first
planet. Ionization fraction may then increase further out in the disk
either via increased midplane temperatures from higher protostellar
heating or via increased received X-ray flux from the protostar or
disk corona \citep[e.g.,][]{2013ApJ...764...65M}.

Following CT14, the predicted formation location of the innermost planet using 
the fiducial Shakura-Sunyaev steady viscous active accretion disk model, and 
assuming negligible protostellar heating, is 
\begin{equation}
\label{eq:rdzib}
r_{1200{\rm K}}=0.178\phidzib\gamma_{1.4}^{-2/9}\kappa_{10}^{2/9}\alpha_{-3}^{-2/9}m_{*,1}^{1/3}(f_r\dot{m}_{-9})^{4/9}\:{\rm AU}.
\end{equation}
Here $\gamma\equiv1.4\gamma_{1.4}$ is the power-law exponent of the
barotropic equation of state $P=K\rho^\gamma$ ($P$ and $\rho$ are
midplane pressure and density),
$\kappa\equiv10\kappa_{10}\:\unitkappa$ is disk opacity,
$\alpha\equiv10^{-3}\alpha_{-3}$
is viscosity parameter,
$m_\star\equiv m_{\star,1}\:\msun$
is stellar mass, $\dot{m}\equiv10^{-9}\dot{m}_{-9}\:\unitmdot$ is
accretion rate, and $f_r\equiv1-\sqrt{r_\star/r}$ where $r_\star$
is stellar radius. Note that the choice of normalization for $\alpha$ 
reflects expected values in the dead-zone region near the DZIB,
and this value is quite uncertain (CT14).
Eq.~\ref{eq:rdzib} is the same as Eq.~11 of CT14 except, we
have added an additional parameter $\phidzib$ to account for the fact
that the estimate of midplane temperature can be affected by
several factors, including energy extraction from a disk wind
and protostellar heating. 
By comparison with more realistic protostellar disk models of 
\citet{2013ApJ...766...86Z}, CT14 argued for a potential reduction in
$r_{1200{\rm{K}}}$ by a factor of two, perhaps also due to reduction in
$\kappa$ as dust grains begin to be destroyed.
Thus for our fiducial model, here we will use $\phidzib=0.5$.

At the location given in Eq.~\ref{eq:rdzib}, a forming planet
may grow in mass, most likely by pebble accretion, to a gap opening
mass determined by the viscous-thermal criterion
\citep{1993prpl.conf..749L},
\begin{eqnarray}\label{eq:Mgap}
M_{p}&=&\frac{\phi_{G}40{\nu}m_*}{r^2\Omega_K}\nonumber\\
&=&20\frac{3^{1/5}}{\pi^{2/5}}\phi_{G}\left(\frac{\mu}{\gamma k_B}\right)^{-4/5}\left(\frac{\kappa}{\sigmasb}\right)^{1/5}\nonumber\\
&\times&\alpha^{4/5}G^{-7/10}m_*^{3/10}\left(f_r\dot{m}\right)^{2/5}r^{1/10}\\
&\rightarrow&5.67\phi_{G,0.3}\gamma_{1.4}^{4/5}\kappa_{10}^{1/5}\alpha_{-3}^{4/5}m_{*,1}^{3/10}(f_r\dot{m}_{-9})^{2/5}\rau^{1/10}\:M_\oplus,\nonumber
\end{eqnarray}
(Eq.~26 of CT14) where we adopt $\phi_{G}=0.3$
\citep{2013ApJ...768..143Z} and $r$ is orbital radius of the forming
planet. An uncertain quantity here is $\dot{m}$, which may vary widely
from system to system and over time within a system.

We eliminate the accretion rate term, $f_r\dot{m}$, from
Eq.~\ref{eq:Mgap} using Eq.~\ref{eq:rdzib} and set $r=r_{\rm 1200K}$
to find the innermost planet mass, $\mpone$, (i.e., gap opening
mass at DZIB) as a function of $r$:
\begin{equation}
\label{eq:mgvsr}
\mpone=5.0\phi_{G,0.3}\phidzib_{0.5}^{-9/10}\gamma_{1.4}\alpha_{-3}(\rau/0.1)\:\mearth,
\end{equation}
i.e., $\mpone\propto\rau$: a linear increase in innermost planet mass
with orbital radius of formation. Note, variation in $r$ is caused by
variation in $\dot{m}$: higher $\dot{m}$ results in $T=1200$~K at
larger radius.  The dependence on $\kappa$ and $m_*$ vanish. The
normalization of the $\mpone$-$r$ relation depends on $\phi_G$,
$\phi_{\rm DZIB}$, $\gamma$ and $\alpha$.

If subsequent planetary migration is negligible, then
Equation~\ref{eq:mgvsr}'s prediction can be compared directly with the
observed STIPs innermost planets.
Two arguments suggest planetary migration from the initial formation
site may be small. First, when the planet is still forming and has not
yet opened a gap, 
drastic change in $\alpha$ at the DZIB creates an outwardly 
increasing surface density gradient and unsaturated corotation torques suppress 
Type I migration \citep{2009MNRAS.394.2283P,2010ApJ...715L..68L} 
creating a ``planet trap" at the (DZIB) pressure maximum 
\citep{2006ApJ...642..478M,2009ApJ...691.1764M}.
Second, when the planet is massive enough to open a gap, its mass
already dominates over that in the inner gas disk, limiting scope for
Type II migration.

\section{Mass, radius, and density of KPCs}
\label{S:obs}
%
%
%
\begin{figure*}
\begin{center}
\plotone{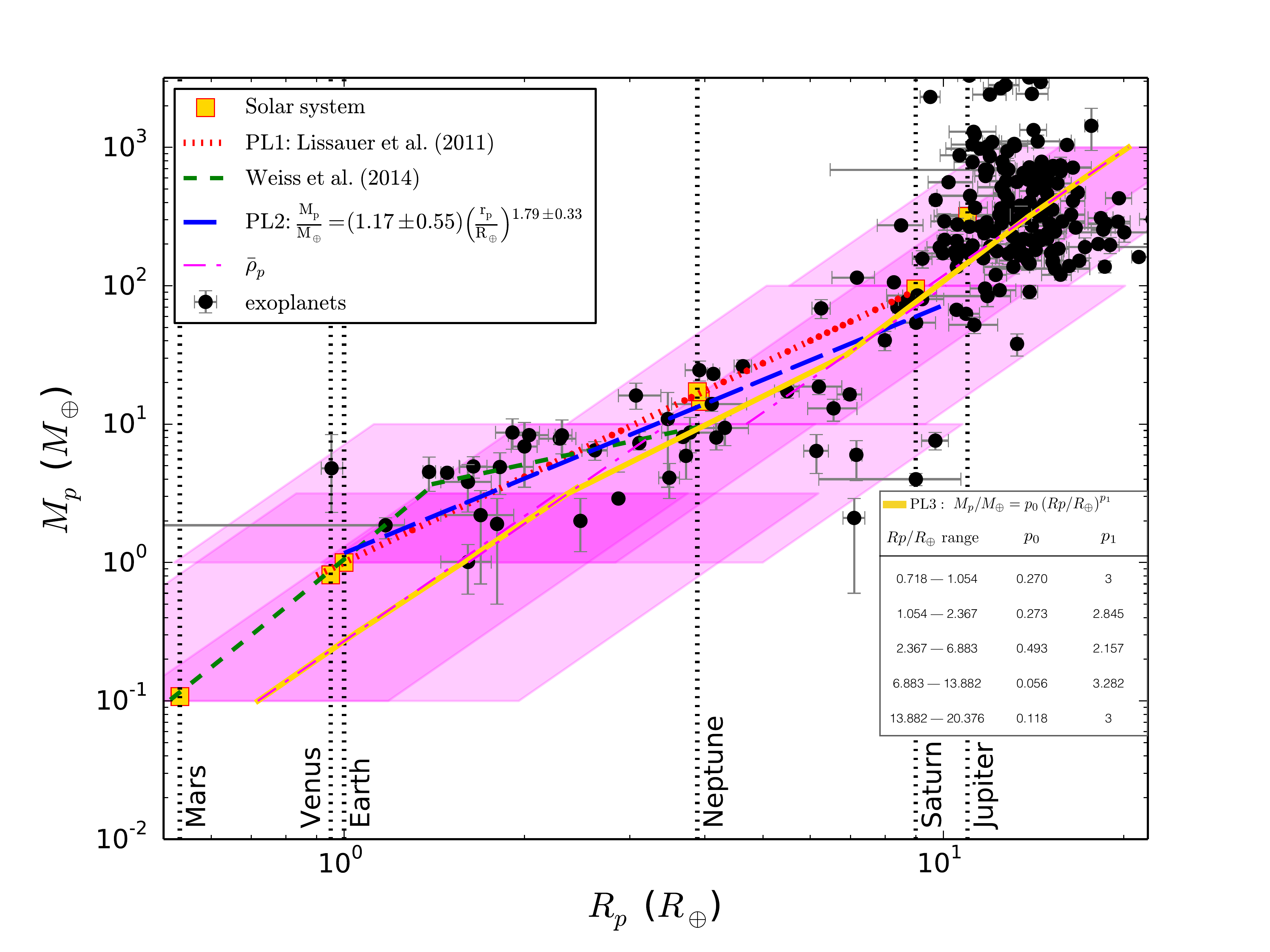}
\caption[$M_p$ vs $R_p$]{
$M_p$ versus $R_p$ for planets with direct measurements of mass and
  size.
Vertical dotted lines and yellow squares show Solar System
planets. 
Black dots denote exoplanets. Dotted (red), dashed (green),
long-dashed (blue), and solid (gold) lines show simple power-law
$M_p$-$R_p$ relations proposed by \citet{2011ApJS..197....8L},
\citet{2014ApJ...783L...6W}, our own best fit power-law
$M_p/\mearth=(1.17\pm0.55)(R_p/\rearth)^{(1.79\pm0.33)}$ over similar
range as \citet{2011ApJS..197....8L}, and piecewise estimated
power-laws based on fitted mean planetary densities in equal intervals
of log~$M_p$ (Fig.~\ref{fig:rho}), respectively. Dash-dot line traces the mean 
$\rho_p$. Darker and fainter shaded regions 
denote $1$ and $2\sigma$ of best-fit lognormal PDFs 
as a function of $M_p$ (Fig.~\ref{fig:rho}).  
}
\label{fig:mprp}
\end{center}
\end{figure*}

While IOPF (CT14) predicts $M_p$, planets of the same mass may attain
widely varying average densities ($\rho_p$) depending on relative
importance of gas and pebble accretion and also atmospheric puffiness,
dependent on detailed atmospheric properties
\citep[e.g.,][]{2014ApJ...787..173H}. Thus predicting $R_p$ is not
straightforward within the framework of IOPF.  However, only $R_p$ is
measured for most \kepler-discovered systems, and especially the
smaller planets exhibit wide ranges of $\rho_p$ even when they are of
comparable sizes
\citep[e.g.,][]{2014ApJ...787..173H,2012ApJ...749...15G,2014ApJ...783...53M}.
Also, both mean and overall range of $\rho_p$ vary based on the planet
mass range under consideration
\citep[e.g.,][Fig.\,\ref{fig:rho}]{2014ApJ...783L...6W,2014ApJ...787..173H}.
Hence, direct comparison between theory and observation is difficult
for individual planets and a statistical approach is needed.

To convert IOPF-predicted $M_p$ into a corresponding $R_p$,
probability distribution functions (PDFs) for $\rho_p$ that change
continuously as a function of $M_p$ would be ideal.  Radial velocity
(RV) follow-up and transit timing variation (TTV) measurements have
constrained $M_p$ for some \kepler\ systems \citep[][for a
  list]{2014ApJS..210...20M}.  However, the small number of observed
planets where $\rho_p$ could be measured limits how finely the
different $M_p$ ranges can be sampled. For this study, we divide the
set of planets with known $\rho_p$ crudely in four groups, each
ranging over 1 dex in $M_p$ with boundaries at $0.1, 1, 10, 10^2$, and
$10^3\:\mearth$. Since no exoplanets with measured $\rho_p$ have
$(M_p/M_\oplus)<1$, we include planets with $M_p$ up to half a dex
into the next group to determine the $\rho_p$-distribution for mass
group $0.1\leq(M_p/\mearth)<1$.
We estimate the observed $\rho_p$ PDFs for each group separately by
fitting lognormals (Fig.~\ref{fig:rho}). We assume that all planets
within each mass group have the same PDF for $\rho_p$. We use the
appropriate $\rho_p$-distribution for an IOPF-predicted $M_p$ for a
given $r$, to randomly generate the average density and calculate
$R_p$ in \S\ref{S:compare}. Note that this division in groups is quite
arbitrary, but necessary given the available data.

$M_p$ values of the thousands of KPCs with measured $R_p$ are often
estimated using simple power-law relations, derived based on planets
with measured $M_p$ 
from RV followup and TTV \citep{2014ApJS..210...20M}.
Although, choosing a simple $M_p$-$R_p$ power-law relation essentially
ignores $\rho_p$ dispersions at a fixed $R_p$, they are popular
because of their simplicity.
Fig.~\ref{fig:mprp} shows a compilation of the data for planets
with directly measured $M_p$ and $R_p$, together with two previously
published fitted $M_p$-$R_p$ power-law relations by
\citet[][henceforth PL1]{2011ApJS..197....8L} and \citet[][]{2014ApJ...783L...6W}.
We also include our own best fit power-law relation following
\citet{2011ApJS..197....8L} for planets between
$1\leq(R_p/\rearth)\leq10$, but not forcing the relation to match the
Earth.  We derive
$(M_p/\mearth)=(1.17\pm0.55)(R_p/\rearth)^{(1.79\pm0.33)}$ (henceforth
PL2) by fitting data with uniform weighting, independent of
measurement errors. This choice is made since we expect the spread in
masses at a given radius reflects an intrinsic dispersion in density
and we wish to avoid the average relation of the planet population
being biased towards the systems that happen to have the smallest
errors. 
Finally, we construct a piecewise power-law 
(henceforth PL3)
by connecting the $R_p$ and $M_p$ values at the middle $R_p$ points in
each $M_p$ group and the mean of $\log\:R_p$ values at the group
boundaries along the mean $\log\rho_p$ lines in each group; 
$M_p/\mearth=p_0(R_p/\rearth)^{p_1}$, $R_p/\rearth$ intervals 
$=\left\{0.718, 1.054, 2.367, 6.883, 13.882, 20.376 \right\}$, 
$p_0=\left\{0.270, 0.273, 0.493, 0.056, 0.118\right\}$, 
$p_1=\left\{3, 2.845, 2.157, 3.282, 3\right\}$. 
%

The estimated $M_p$ can thus be different for the same observed $R_p$
depending on which power-law is used. However, Fig.~\ref{fig:mprp}
shows that the intrinsic dispersion in $M_p$ at a given $R_p$, due to
a dispersion in $\rho_p$, is larger than the differences between the
power laws. For completeness we will use all three power-laws PL1--3
to estimate $M_p$ for a given $R_p$ and show the resulting
differences. For this study we do not use the power-law proposed by
\citet{2014ApJ...783L...6W} since its applicability is within a
limited range in $R_p\leq4\,\rearth$.

\section{Comparison with observed Kepler systems}
\label{S:compare}
%
%
%
\begin{figure*}
\begin{center}
\plotone{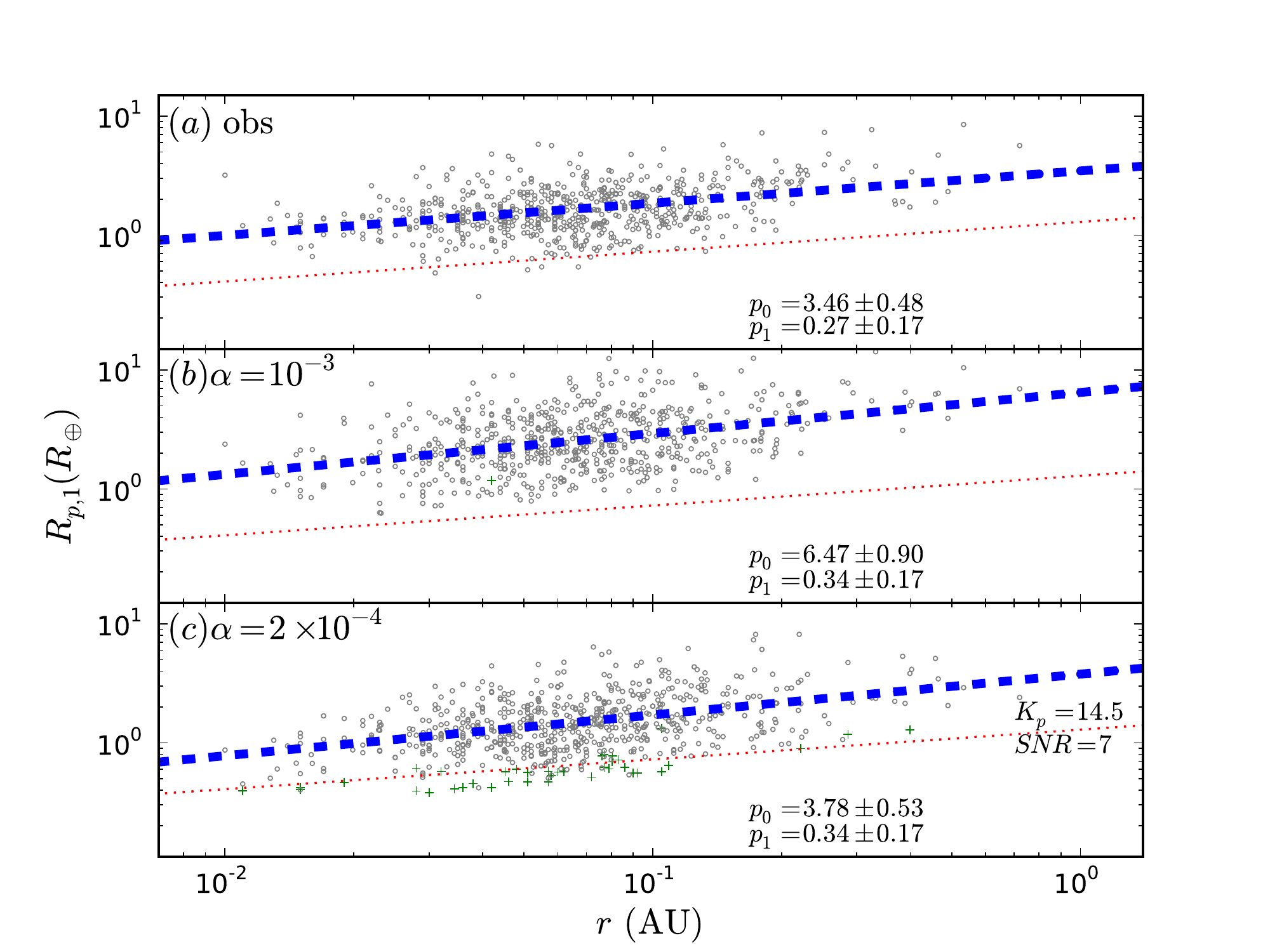}
\caption[$\rpone$ vs $r$]{
Planetary radius of innermost planets ($\rpone$) in multiplanet
systems as a function of $r$ (grey dots). Green '+'s denote planets
discarded because they would not be detectable by \kepler. Blue dashed
line represents best-fit power-law $\rpone/\rearth=p_0\rau^{p_1}$. Red
dotted lines show detection limit (SNR=7) for median $K_p=14.5$ of
host stars in the observed sample. (a) Top: Observed
\kepler\ population. (b) Middle: Synthetically generated planet
population from IOPF (\S\ref{S:compare}) using $\alpha=10^{-3}$. (c)
Bottom: As (b), but for $\alpha=2\times10^{-4}$. Best-fit values of
$p_0$ and $p_1$ are shown in each panel, including $1\sigma$ errors.
Within statistical fluctuations, $p_1$ values for the synthetic
populations (both $\alpha$s) agree well with the observed scaling.
}
\label{fig:rpvsa}
\end{center}
\end{figure*} 
%
%
%
%
%
\begin{figure*}
\begin{center}
\plotone{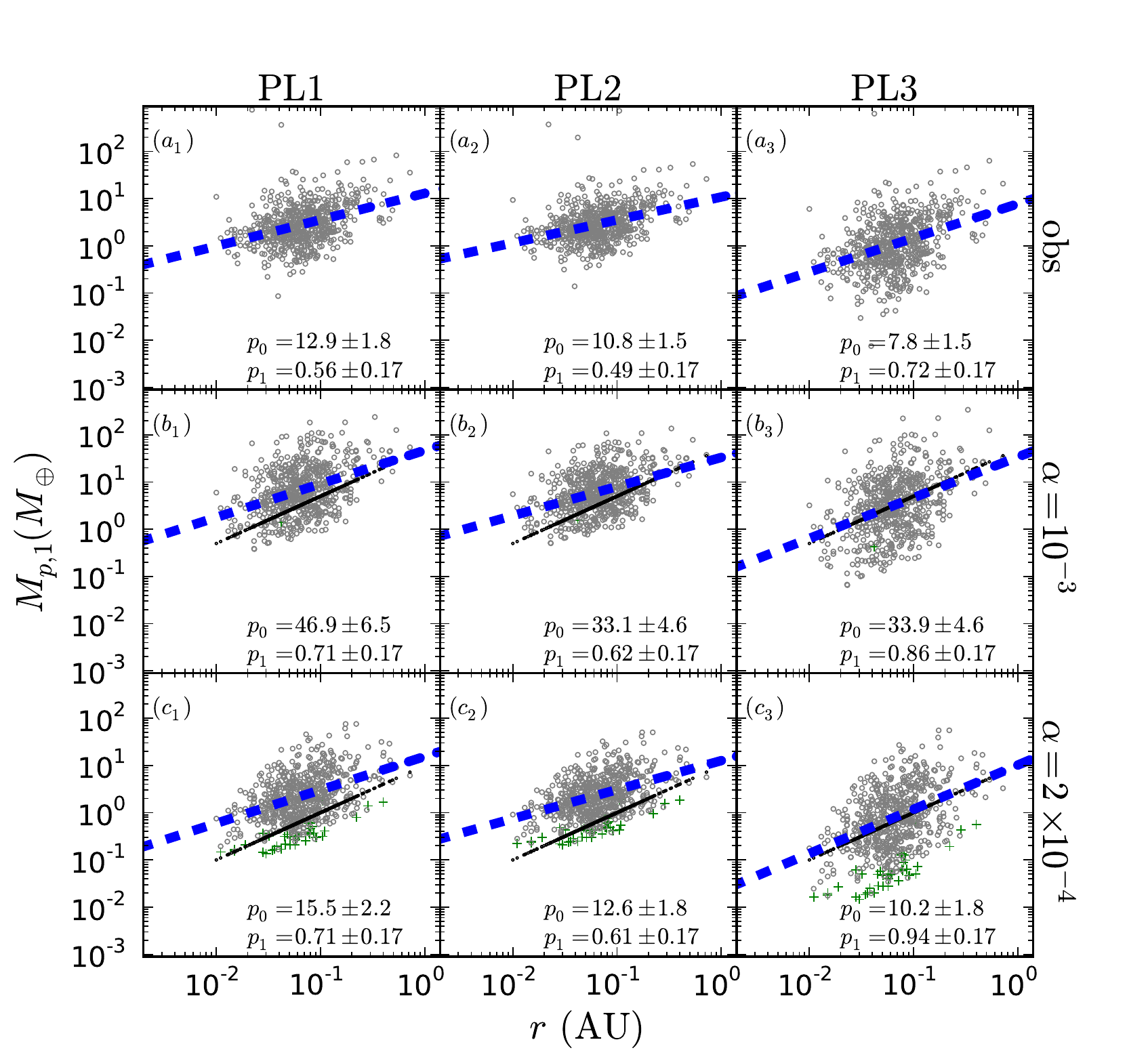}
\caption[$\mpone$ vs $r$]{
Mass of innermost planets ($\mpone$) in multiplanet systems versus
$r$. Left to right, panels show $\mpone$ values obtained using PL1--3,
respectively (\S\ref{S:obs}).  Top to bottom, panels show observed and
synthetic data with $\alpha=10^{-3}$ and $2\times10^{-4}$,
respectively (\S\ref{S:compare}).
Black dots denote the actual $\mpone$ of
the synthetic data, following Eq.~\ref{eq:mgvsr}. 
Grey dots denote estimated $\mpone$ for a given $R_p$ using one of the
$M_p$-$R_p$ power-laws (PL1--3). Green '+'s are
undetectable planets. Blue dashed lines show best-fit power-laws
$\mpone/\mearth=p_0\rau^{p_1}$, $p_0$ and $p_1$ with $1\sigma$
errors shown in each panel. PL1,2 systematically predict a higher $\rho_p$ 
for a given $R_p$ relative to our fitted lognormal mean $\rho_p$ (Fig.~\ref{fig:mprp}) resulting in typically 
higher estimated masses compared to actual $M_p$. 
}
\label{fig:mpvsa}
\end{center}
\end{figure*} 

Since we are interested in testing whether properties of STIPs
innermost planets are consistent with IOPF predictions, we restrict
ourselves only to innermost KPCs in multiplanet systems
($n_p\geq2$).
We obtain KPC data from NASA's exoplanet archive
(http://exoplanetarchive.ipac.caltech.edu; June 25, 2014 update).  We
find that for the 629 multi-transiting systems,
$\rpone/\rearth=(3.5\pm0.5)\rau^{(0.3\pm0.2)}$, where errors are
$1\sigma$ obtained from parameter estimation and fitting is done using
equal weight to each data point (Fig.~\ref{fig:rpvsa}a).

While creating the synthetic innermost planet populations based on the
IOPF model we pay attention to replicate all observational biases in
the observed sample as closely as possible.
We import the period $P$, semimajor axis $a$, assumed to be equal to 
$r$ (low eccentricity), $r_\star$, and \kepler\ magnitude ($K_p$) for 
the innermost KPCs. 
This way our synthetic planet sample automatically preserves the
observed distribution of planetary orbital and host star properties.
For a given $r$ we use Eq.~\ref{eq:mgvsr} to determine $M_p$ as
predicted by IOPF. Densities are then randomly assigned by drawing
from the appropriate lognormal PDFs (\S\ref{S:obs}).
We restrict $\rho_p$ to be between $32$ and $0.01\:\gccc$
\citep{2014ApJ...787..173H,2014ApJ...783...53M}. Our conclusions are
not very sensitive to reasonable changes in the $\rho_p$ range.
Note, the actual total range in $\rho_p$ is unknown and transit
observations are biased towards detecting lower density planets in
general. Planet size $R_p$ is calculated using $M_p$ and $\rho_p$.
Using host star $K_p$ values we estimate the combined differential
photometric precision following 
\citet[][see \citealt{2012MNRAS.427.1587C} for details]{2011ApJS..197....6G}. We then
estimate whether this synthetic planet would be detectable (SNR$>7$
assuming $3.5\yr$ observation) by \kepler. We repeat this process
until we generate one \kepler-detectable planet for each host
star. Examples of synthetic populations, each of $629$ detectable planets, are
shown in Fig.~\ref{fig:rpvsa}(b) using $\alpha=10^{-3}$, and (c) using
$\alpha=2\times10^{-4}$.

We find IOPF-predicted $\rpone$ versus $r$,
$\rpone/\rearth\propto\rau^{0.3\pm0.2}$, shows very similar scaling as
that of the observed planets.
The absolute normalization is somewhat arbitrary and depends on
unconstrained disk properties including $\alpha$ and $\phidzib$. For
example, while the scaling remains very similar, the proportionality
constant changes with a change of the adopted value of $\alpha$. For
$\alpha=2\times10^{-4}$ both the scaling and the normalization agree
well for the $\rpone$--$r$ relations in the observed and synthetic
samples.

Turning to masses, the 
$\mpone$-$r$ relation depends on the adopted
$M_p$-$R_p$ relation. For PL1--3 these are given as
$\mpone=(12.9\pm1.8)\rau^{(0.56\pm0.17)},\:(10.8\pm1.5)\rau^{(0.49\pm0.17)},\:(7.8\pm1.5)=\rau^{(0.72\pm0.17)}$,
respectively for the observed sample (Fig.~\ref{fig:mpvsa}).  Thus,
adopting a simple $M_p$-$R_p$ relation, or equivalently, assuming a
fixed $\rho_p$ for a given $R_p$ in estimating $M_p$ results in
$\mpone$-$r$ scalings that are shallower than the linear prediction of
IOPF (Eq.~\ref{eq:mgvsr}).

Fig.~\ref{fig:mpvsa} shows the comparison between observed and
synthetic populations for PL1--3 and for $\alpha=10^{-3}$ and
$2\times10^{-4}$. We find that 
for all considered simple $M_p$-$R_p$ relations (PL1--3), best-fit
power laws for observed and predicted planet populations agree
reasonably well. As for the $\rpone$-$r$ relations, the
scalings agree within expected statistical fluctuations 
for both $\alpha$ values. The normalization is again off by a factor
of a few for $\alpha=10^{-3}$, but is quite similar for
$\alpha=2\times10^{-4}$ for all $M_p$-$R_p$ power-laws. 
It is also instructive to see the degree to which estimated $M_p$ can
diverge from actual $M_p$ due to the assumption of fixed $\rho_p$ for
fixed $R_p$, or equivalently, assuming a simple power-law relation
between $R_p$ and $M_p$. Using such power-laws, while useful for a
crude estimate of $M_p$ from an observed $R_p$, can lead to derived
$M_p$ being very different from the actual one, due to the intrinsic
dispersion in density.
This highlights the importance of further TTV analysis and RV
followup.

%
%
%
%
\begin{figure*}
\begin{center}
\plotone{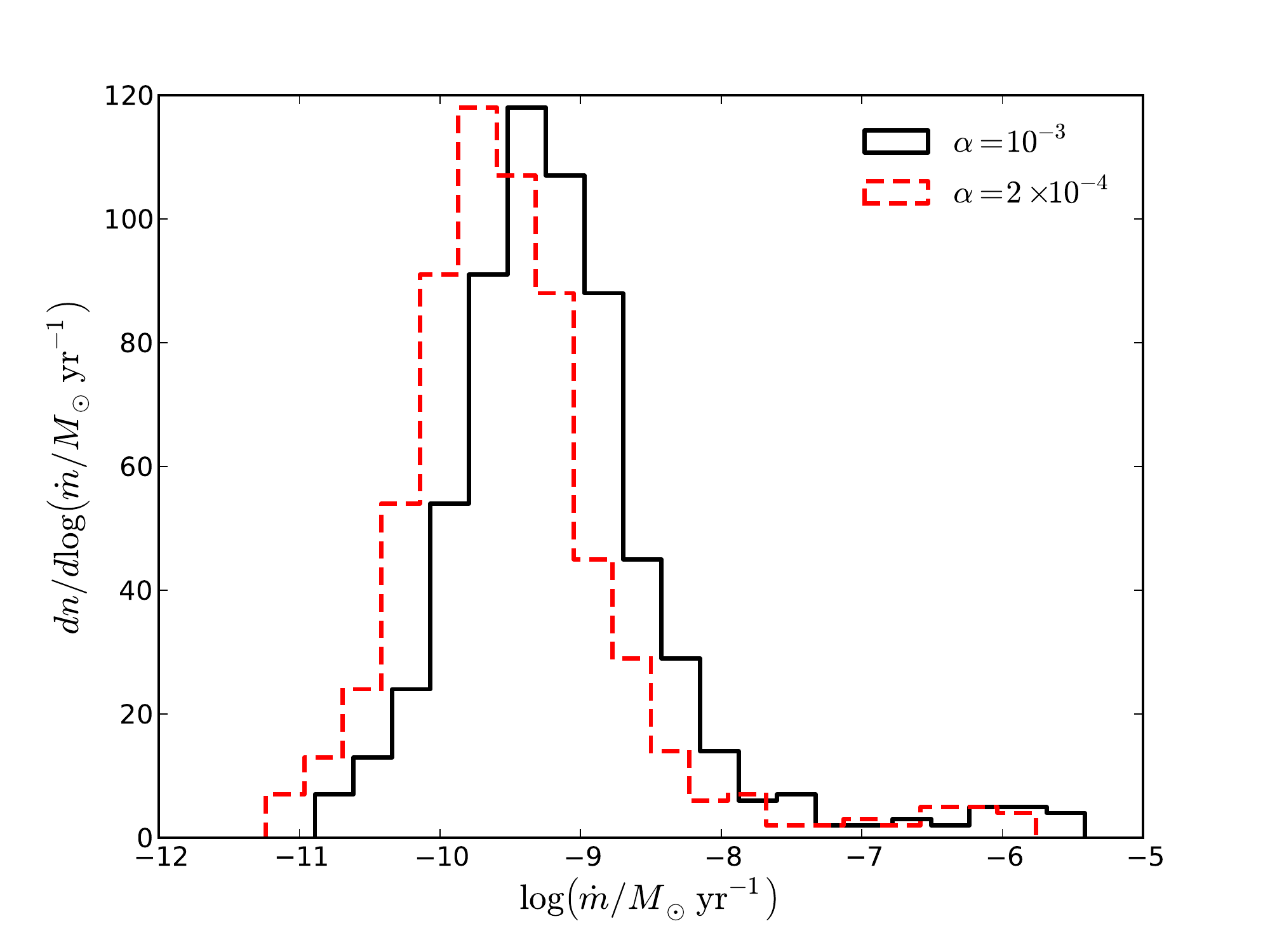}
\caption[$\dot{m_\star}$ distribution]{
Histogram of effective mass accretion rate ($\dot{m}$) for creation of
the innermost observed planets in \kepler's multitransiting
systems. The $\dot{m}$ values are calculated using fiducial values
described in \S\ref{S:derivation} and $\alpha=10^{-3}$
(black solid) and $2\times10^{-4}$ (red dashed). The innermost planet 
candidates are chosen as described in
\S\ref{S:compare}. 
For $\alpha=2\times10^{-4}$ the $16$, $50$, and $84$th
percentiles are $7.2\times10^{-11}$, $2.6\times10^{-10}$, and
$1.2\times10^{-9}\:\smyr$, respectively.
}
\label{fig:mdot}
\end{center}
\end{figure*} 
Assuming that our selected observed sample of innermost planets truly
are innermost, ``Vulcan'' planets, their observed orbital radii can
also constrain $\dot{m}$ via Eq.~\ref{eq:rdzib}. Figure~\ref{fig:mdot}
shows histograms of the expected $\dot{m}$ for the observed systems if
formed via IOPF. The estimated effective $\dot{m}$ for a given
innermost planet's position depends on $\alpha$. We find that the
majority of the observed sample of innermost planets predict effective
$\dot{m}$ between $\sim10^{-11}$--$10^{-8}\:\unitmdot$ for
$\alpha=2\times10^{-4}$.  The tail towards very large
$\dot{m}\gtrsim10^{-7}$ may indicate that some selected planets are
not actually innermost planets: either there is an undetected inner
planet
\citep{2012Sci...336.1133N,2013ApJ...777....3N,2014A&A...561L...1B},
or perhaps the original inner planet has been removed via, for
example, collision or ejection.

\section{Discussion and Conclusions}
\label{S:conclude}
We showed that IOPF predicts STIPs innermost planet mass, $\mpone$,
increases linearly with $r$, independent of $\dot{m}$, $m_*$, or
$\kappa$. Absolute values for $\mpone$, however, depend strongly on
disk properties, especially viscosity parameter $\alpha$.

Using fiducial disk parameters and observationally motivated
mass-based density ranges we found the IOPF $\rpone$-$r$ scaling is
consistent with that in observed \kepler\ multis
(Fig.~\ref{fig:rpvsa}). Comparing mass scalings involved assuming a
$M_p$-$R_p$ relation (Fig.~\ref{fig:mprp}).
The estimated $\mpone$-$r$ scalings vary depending on which
$M_p$-$R_p$ relation was chosen, even when the real underlying
relation is $\mpone\propto{r}$. We showed that $\mpone$-$r$ scalings
for theoretical and observed populations
agree within expected uncertainties for all adopted $M_p$-$R_p$
relations. Assuming formation via IOPF, the distribution of $r$ for
the innermost planets implies $\dot{m}$ between
$\sim10^{-10}$--$10^{-9}\:\unitmdot$, adopting our preferred DZIB
$\alpha=2\times10^{-4}$.

For comparison between the IOPF predicted and observed inner planet
properties we had to make several simplifying assumptions. We assumed
the $M_p$-based $\rho_p$ distributions for STIPs innermost planets are
similar to those obtained from all planets with known $\rho_p$.
However, IOPF innermost planets, forming very close to the star,
potentially have quite different entropy structure in their
atmospheres and thus systematically different densities compared to
planets that form further out. We have also assumed that the apparent
innermost planet in multitransiting systems are truly so. Limiting the
observed sample to only multitransiting systems and only out to
$r=1\:\au$ alleviates this problem somewhat.

The exact $\rpone$-$r$ and $\mpone$-$r$ relations for the synthetic
planet population predicted by IOPF depend somewhat on the $M_p$-based
$\rho_p$ distributions adopted. However, for several observationally
motivated $\rho_p$ distributions we find agreement between the
theoretical and observed populations.
Nevertheless, since the $\rho_p$ PDF in the lowest-mass bin is the
most important in determining the fraction of detectable small planets
in a synthetic population, more observational constraints on densities
of low-mass planets ($M_p<1\mearth$) will be very useful for a more
robust comparison. Continued efforts for RV followup and TTV
measurements will potentially lead to more mass measurements making a
more direct comparison possible for testing IOPF theory.
Another source of change in the final $\mpone$ and $\rpone$ vs $r$
relations is possible inward migration of some planets after they have
formed via IOPF, which needs to be investigated in future numerical
simulations. 

Finally, 
we point out that there may be other mechanisms that can help create
$M_{p}$ vs $r$ correlations. For example,
increasing planet mass with radius may be related to
radial dependence of massloss via stellar irradiation
\citep{2012ApJ...761...59L}. A quantitative calculation of the effects
of this mass-loss, which will tend to steepen the IOPF $\mpone$ vs $r$
relation, will require modeling the composition and size of the planets,
as well as the history of their EUV and X-ray flux exposure.

\acknowledgements This research used the NASA Exoplanet Archive
operated by Caltech, under contract with NASA (Exoplanet Exploration
Program). SC acknowledges support from UF Theory and
CIERA-Northwestern postdoctoral fellowships.

\begin{thebibliography}{43}
\expandafter\ifx\csname natexlab\endcsname\relax\def\natexlab#1{#1}\fi

\bibitem[{{Barros} {et~al.}(2014){Barros}, {D{\'{\i}}az}, {Santerne}, {Bruno},
  {Deleuil}, {Almenara}, {Bonomo}, {Bouchy}, {Damiani}, {H{\'e}brard},
  {Montagnier}, \& {Moutou}}]{2014A&A...561L...1B}
{Barros}, S.~C.~C., {D{\'{\i}}az}, R.~F., {Santerne}, A., {Bruno}, G.,
  et~al. 2014, \aap, 561, L1

\bibitem[{{Baruteau} \& {Papaloizou}(2013)}]{2013ApJ...778....7B}
{Baruteau}, C. \& {Papaloizou}, J.~C.~B. 2013, \apj, 778, 7

\bibitem[{{Batygin} \& {Morbidelli}(2013)}]{2013AJ....145....1B}
{Batygin}, K. \& {Morbidelli}, A. 2013, \aj, 145, 1

\bibitem[{{Chatterjee} \& {Ford}(2014)}]{2014arXiv1406.0521C}
{Chatterjee}, S. \& {Ford}, E.~B. 2014, arXiv:1406.0512

\bibitem[{{Chatterjee} {et~al.}(2012){Chatterjee}, {Ford}, {Geller}, \&
  {Rasio}}]{2012MNRAS.427.1587C}
{Chatterjee}, S., {Ford}, E.~B., {Geller}, A.~M., \& {Rasio}, F.~A. 2012,
  \mnras, 427, 1587

\bibitem[{{Chatterjee} {et~al.}(2008){Chatterjee}, {Ford}, {Matsumura}, \&
  {Rasio}}]{2008ApJ...686..580C}
{Chatterjee}, S., {Ford}, E.~B., {Matsumura}, S., \& {Rasio}, F.~A. 2008, \apj,
  686, 580

\bibitem[{{Chatterjee} \& {Tan}(2014)}]{2014ApJ...780...53C}
{Chatterjee}, S. \& {Tan}, J.~C. 2014, \apj, 780, 53

\bibitem[{{Chiang} \& {Laughlin}(2013)}]{2013MNRAS.431.3444C}
{Chiang}, E. \& {Laughlin}, G. 2013, \mnras, 431, 3444

\bibitem[{{Cossou} {et~al.}(2014){Cossou}, {Raymond}, {Hersant}, \&
  {Pierens}}]{2014arXiv1407.6011C}
{Cossou}, C., {Raymond}, S.~N., {Hersant}, F., \& {Pierens}, A. 2014,
  arXiv:1407.6011

\bibitem[{{Cossou} {et~al.}(2013){Cossou}, {Raymond}, \&
  {Pierens}}]{2013A&A...553L...2C}
{Cossou}, C., {Raymond}, S.~N., \& {Pierens}, A. 2013, \aap, 553, L2

\bibitem[{{Fang} \& {Margot}(2012)}]{2012ApJ...761...92F}
{Fang}, J. \& {Margot}, J.-L. 2012, \apj, 761, 92

\bibitem[{{Gautier} {et~al.}(2012){Gautier}, {Charbonneau}, {Rowe}, {Marcy},
  {Isaacson}, {Torres}, {Fressin}, {Rogers}, {D{\'e}sert}, {Buchhave},
  {Latham}, {Quinn}, {Ciardi}, {Fabrycky}, {Ford}, {Gilliland}, {Walkowicz},
  {Bryson}, {Cochran}, {Endl}, {Fischer}, {Howell}, {Horch}, {Barclay},
  {Batalha}, {Borucki}, {Christiansen}, {Geary}, {Henze}, {Holman}, {Ibrahim},
  {Jenkins}, {Kinemuchi}, {Koch}, {Lissauer}, {Sanderfer}, {Sasselov},
  {Seager}, {Silverio}, {Smith}, {Still}, {Stumpe}, {Tenenbaum}, \& {Van
  Cleve}}]{2012ApJ...749...15G}
{Gautier}, III, T.~N., {Charbonneau}, D., {Rowe}, J.~F., {Marcy}, G.~W.,
  et~al. 2012, \apj, 749, 15

\bibitem[{{Gilliland} {et~al.}(2011){Gilliland}, {Chaplin}, {Dunham},
  {Argabright}, {Borucki}, {Basri}, {Bryson}, {Buzasi}, {Caldwell}, {Elsworth},
  {Jenkins}, {Koch}, {Kolodziejczak}, {Miglio}, {van Cleve}, {Walkowicz}, \&
  {Welsh}}]{2011ApJS..197....6G}
{Gilliland}, R.~L., {Chaplin}, W.~J., {Dunham}, E.~W., {Argabright}, V.~S., 
et~al. 2011, \apjs, 197, 6

\bibitem[{{Goldreich} \& {Schlichting}(2014)}]{2014AJ....147...32G}
{Goldreich}, P. \& {Schlichting}, H.~E. 2014, \aj, 147, 32

\bibitem[{{Hansen} \& {Murray}(2012)}]{2012ApJ...751..158H}
{Hansen}, B.~M.~S. \& {Murray}, N. 2012, \apj, 751, 158

\bibitem[{{Hansen} \& {Murray}(2013)}]{2013ApJ...775...53H}
---. 2013, \apj, 775, 53

\bibitem[{{Howe} {et~al.}(2014){Howe}, {Burrows}, \&
  {Verne}}]{2014ApJ...787..173H}
{Howe}, A.~R., {Burrows}, A., \& {Verne}, W. 2014, \apj, 787, 173

\bibitem[{{Hu} {et~al.}(2014){Hu}, {Tan}, \&
  {Chatterjee}}]{2014arXiv1410.5819H}
{Hu}, X., {Tan}, J.~C., \& {Chatterjee}, S. 2014, arXiv:1410.5819

\bibitem[{{Kley} \& {Nelson}(2012)}]{2012ARA&A..50..211K}
{Kley}, W. \& {Nelson}, R.~P. 2012, \araa, 50, 211

\bibitem[{{Lin} \& {Papaloizou}(1993)}]{1993prpl.conf..749L}
{Lin}, D.~N.~C. \& {Papaloizou}, J.~C.~B. 1993, in Protostars and Planets III,
  ed. E.~H. {Levy} \& J.~I. {Lunine}, 749--835

\bibitem[{{Lissauer} {et~al.}(2011){Lissauer}, {Ragozzine}, {Fabrycky},
  {Steffen}, {Ford}, {Jenkins}, {Shporer}, {Holman}, {Rowe}, {Quintana},
  {Batalha}, {Borucki}, {Bryson}, {Caldwell}, {Carter}, {Ciardi}, {Dunham},
  {Fortney}, {Gautier}, {Howell}, {Koch}, {Latham}, {Marcy}, {Morehead}, \&
  {Sasselov}}]{2011ApJS..197....8L}
{Lissauer}, J.~J., {Ragozzine}, D., {Fabrycky}, D.~C., {Steffen}, J.~H.,
  et~al. 2011, \apjs, 197, 8

\bibitem[{{Lithwick} \& {Wu}(2012)}]{2012ApJ...756L..11L}
{Lithwick}, Y. \& {Wu}, Y. 2012, \apjl, 756, L11

\bibitem[{{Lopez} {et~al.}(2012){Lopez}, {Fortney}, \&
  {Miller}}]{2012ApJ...761...59L}
{Lopez}, E.~D., {Fortney}, J.~J., \& {Miller}, N. 2012, \apj, 761, 59

\bibitem[{{Lyra} {et~al.}(2010){Lyra}, {Paardekooper}, \& {Mac
  Low}}]{2010ApJ...715L..68L}
{Lyra}, W., {Paardekooper}, S.-J., \& {Mac Low}, M.-M. 2010, \apjl, 715, L68

\bibitem[{{Marcy} {et~al.}(2014){Marcy}, {Isaacson}, {Howard}, {Rowe},
  {Jenkins}, {Bryson}, {Latham}, {Howell}, {Gautier}, {Batalha}, {Rogers},
  {Ciardi}, {Fischer}, {Gilliland}, {Kjeldsen}, {Christensen-Dalsgaard},
  {Huber}, {Chaplin}, {Basu}, {Buchhave}, {Quinn}, {Borucki}, {Koch}, {Hunter},
  {Caldwell}, {Van Cleve}, {Kolbl}, {Weiss}, {Petigura}, {Seager}, {Morton},
  {Johnson}, {Ballard}, {Burke}, {Cochran}, {Endl}, {MacQueen}, {Everett},
  {Lissauer}, {Ford}, {Torres}, {Fressin}, {Brown}, {Steffen}, {Charbonneau},
  {Basri}, {Sasselov}, {Winn}, {Sanchis-Ojeda}, {Christiansen}, {Adams},
  {Henze}, {Dupree}, {Fabrycky}, {Fortney}, {Tarter}, {Holman}, {Tenenbaum},
  {Shporer}, {Lucas}, {Welsh}, {Orosz}, {Bedding}, {Campante}, {Davies},
  {Elsworth}, {Handberg}, {Hekker}, {Karoff}, {Kawaler}, {Lund}, {Lundkvist},
  {Metcalfe}, {Miglio}, {Silva Aguirre}, {Stello}, {White}, {Boss}, {Devore},
  {Gould}, {Prsa}, {Agol}, {Barclay}, {Coughlin}, {Brugamyer}, {Mullally},
  {Quintana}, {Still}, {Thompson}, {Morrison}, {Twicken}, {D{\'e}sert},
  {Carter}, {Crepp}, {H{\'e}brard}, {Santerne}, {Moutou}, {Sobeck}, {Hudgins},
  {Haas}, {Robertson}, {Lillo-Box}, \& {Barrado}}]{2014ApJS..210...20M}
{Marcy}, G.~W., {Isaacson}, H., {Howard}, A.~W., {Rowe}, J.~F., 
et~al. 2014, \apjs, 210, 20

\bibitem[{{Masset} {et~al.}(2006){Masset}, {Morbidelli}, {Crida}, \&
  {Ferreira}}]{2006ApJ...642..478M}
{Masset}, F.~S., {Morbidelli}, A., {Crida}, A., \& {Ferreira}, J. 2006, \apj,
  642, 478

\bibitem[{{Masuda}(2014)}]{2014ApJ...783...53M}
{Masuda}, K. 2014, \apj, 783, 53

\bibitem[{{Matsumoto} {et~al.}(2012){Matsumoto}, {Nagasawa}, \&
  {Ida}}]{2012Icar..221..624M}
{Matsumoto}, Y., {Nagasawa}, M., \& {Ida}, S. 2012, \icarus, 221, 624

\bibitem[{{Matsumura} {et~al.}(2009){Matsumura}, {Pudritz}, \&
  {Thommes}}]{2009ApJ...691.1764M}
{Matsumura}, S., {Pudritz}, R.~E., \& {Thommes}, E.~W. 2009, \apj, 691, 1764

\bibitem[{{Mohanty} {et~al.}(2013){Mohanty}, {Ercolano}, \&
  {Turner}}]{2013ApJ...764...65M}
{Mohanty}, S., {Ercolano}, B., \& {Turner}, N.~J. 2013, \apj, 764, 65

\bibitem[{{Nagasawa} \& {Ida}(2011)}]{2011ApJ...742...72N}
{Nagasawa}, M. \& {Ida}, S. 2011, \apj, 742, 72

\bibitem[{{Nesvorn{\'y}} {et~al.}(2013){Nesvorn{\'y}}, {Kipping}, {Terrell},
  {Hartman}, {Bakos}, \& {Buchhave}}]{2013ApJ...777....3N}
{Nesvorn{\'y}}, D., {Kipping}, D., {Terrell}, D., {Hartman}, J., {Bakos},
  G.~{\'A}., \& {Buchhave}, L.~A. 2013, \apj, 777, 3

\bibitem[{{Nesvorn{\'y}} {et~al.}(2012){Nesvorn{\'y}}, {Kipping}, {Buchhave},
  {Bakos}, {Hartman}, \& {Schmitt}}]{2012Sci...336.1133N}
{Nesvorn{\'y}}, D., {Kipping}, D.~M., {Buchhave}, L.~A., {Bakos}, G.~{\'A}.,
  {Hartman}, J., \& {Schmitt}, A.~R. 2012, Science, 336, 1133

\bibitem[{{Paardekooper} \& {Papaloizou}(2009)}]{2009MNRAS.394.2283P}
{Paardekooper}, S.-J. \& {Papaloizou}, J.~C.~B. 2009, \mnras, 394, 2283

\bibitem[{{Papaloizou}(2011)}]{2011CeMDA.111...83P}
{Papaloizou}, J.~C.~B. 2011, Celestial Mechanics and Dynamical Astronomy, 111,
  83

\bibitem[{{Petrovich} {et~al.}(2013){Petrovich}, {Malhotra}, \&
  {Tremaine}}]{2013ApJ...770...24P}
{Petrovich}, C., {Malhotra}, R., \& {Tremaine}, S. 2013, \apj, 770, 24

\bibitem[{{Rasio} \& {Ford}(1996)}]{1996Sci...274..954R}
{Rasio}, F.~A. \& {Ford}, E.~B. 1996, Science, 274, 954

\bibitem[{{Raymond} \& {Cossou}(2014)}]{2014MNRAS.440L..11R}
{Raymond}, S.~N. \& {Cossou}, C. 2014, \mnras, 440, L11

\bibitem[{{Rein}(2012)}]{2012MNRAS.427L..21R}
{Rein}, H. 2012, \mnras, 427, L21

\bibitem[{{Schlichting}(2014)}]{2014ApJ...795L..15S}
{Schlichting}, H.~E. 2014, \apjl, 795, L15

\bibitem[{{Weiss} \& {Marcy}(2014)}]{2014ApJ...783L...6W}
{Weiss}, L.~M. \& {Marcy}, G.~W. 2014, \apjl, 783, L6

\bibitem[{{Zhang} {et~al.}(2013){Zhang}, {Tan}, \&
  {McKee}}]{2013ApJ...766...86Z}
{Zhang}, Y., {Tan}, J.~C., \& {McKee}, C.~F. 2013, \apj, 766, 86

\bibitem[{{Zhu} {et~al.}(2013){Zhu}, {Stone}, \&
  {Rafikov}}]{2013ApJ...768..143Z}
{Zhu}, Z., {Stone}, J.~M., \& {Rafikov}, R.~R. 2013, \apj, 768, 143

\end{thebibliography}
%

%

%
\end{document}